\documentclass[cits]{PoS}

\title{The chiral partner of the nucleon in the mirror assignment with global symmetry}

\ShortTitle{The chiral partner of the nucleon}

\author{\speaker{Susanna Gallas}%
\\
        Johann Wolfgang Goethe-Universit\"{a}t Frankfurt am Main\\
        E-mail: \email{gallas@th.physik.uni-frankfurt.de}}

\author{Francesco Giacosa\\
        Johann Wolfgang Goethe-Universit\"{a}t Frankfurt am Main\\
        E-mail: \email{giacosa@th.physik.uni-frankfurt.de}}

\author{Dirk H. Rischke\\
        Johann Wolfgang Goethe-Universit\"{a}t Frankfurt am Main\\
        E-mail: \email{drischke@th.physik.uni-frankfurt.de}}

\abstract{We calculate the pion-nucleon scattering lengths $a_{0}^{(\pm)}$ and the mass parameter $m_{0}$, which describes the nucleon mass in the chiral limit, at tree-level in the framework of a globally symmetric linear sigma model with parity-doubled nucleons. When recent lattice results \cite{Takahashi} are used, we obtain $m_{0}\simeq300-600$ MeV. While $a_{0}^{(-)}$ is in fair agreement with experimental data, $a_{0}^{(+)}$ is too small because of the employed large scalar meson mass. This indicates the need to account for additional scalar degrees of freedom.} 

\FullConference{8th Conference Quark Confinement and the Hadron Spectrum \\
		 September 1-6 2008\\
		 Mainz, Germany}

\begin{document}

\section{Introduction}

Effective models which embody chiral symmetry and its spontaneous breakdown
at low temperatures and densities are widely used to understand the
properties of light hadrons. Viable candidates obey a well-defined set of
low-energy theorems \cite{meissner,gasioro} but they still differ in
some crucial and interesting aspects such as the mass generation of the
nucleon and the behavior at non-zero $T$ and $\mu $.

Here we concentrate on a linear sigma model with $U(2)_{R}\times U(2)_{L}$
symmetry and parity-doubled nucleons. The mesonic sector involves scalar,
pseudoscalar, vector, and axial-vector mesons. In the baryonic sector,
besides the usual nucleon doublet field $N,$ a second baryon doublet $%
N^{\ast }$ with $J^{P}=\frac{1}{2}^{-}$ is included. As first discussed in
Ref.\ \cite{DeTar:1988kn} and extensively analyzed in Ref.\ \cite{jido}, in
the so-called mirror assignment the nucleon fields $N$ and $N^{\ast }$ have a mass 
$m_{0}\neq 0$ in the chirally symmetric phase. The chiral condensate $\varphi $ increases the masses and generates a mass splitting of $N$ and $N^{\ast },$ but is no
longer solely responsible for generating the masses. Such a theoretical set-up has been used in Ref.\ \cite%
{zschiesche} to study the properties of cold and dense nuclear matter. The
experimental assignment for the chiral partner of the nucleon is still
controversial: the well-identified resonances $N^{\ast }(1535)$ and $N^{\ast
}(1650)$ are two candidates with the right quantum numbers listed in the PDG 
\cite{PDG}, but we shall also investigate the possibility of a very broad
and not yet discovered resonance centered at about $1.2$ GeV, which has been
proposed in Ref.\ \cite{zschiesche}.

The aim of the present work is the development of an effective model which
embodies the chiral partner of the nucleon in the mirror assignment within
the context of global chiral symmetry involving also vector and axial-vector
mesons. In this way more terms appear than those originally proposed in Ref.\ \cite%
{DeTar:1988kn}. Moreover, we restrict our study to operators up to fourth order (thus
not including Weinberg-Tomozawa interaction terms). The axial coupling
constant of the nucleon can be correctly described. Using recent information
about the axial coupling of the partner \cite{Takahashi} and experimental
knowledge about its decay width \cite{PDG}\ we can evaluate
the mass parameter $m_{0}$ which describes the nucleon mass in the chiral limit.
 Then, we further evaluate pion-nucleon scattering lengths
and we compare them with the experimental values \cite{schroder}.
\newline

\section{The model and its implications}

The scalar and pseudoscalar fields are included in the matrix $\Phi =(\sigma
+i\eta )t^{0}+(\overrightarrow{a}_{0}+i\overrightarrow{\pi })\cdot\overrightarrow{%
t}$ and the (axial-)vector fields are represented by the matrices $%
R^{\mu }=(\omega ^{\mu }-f_{1}^{\mu })t^{0}+(\overrightarrow{\rho }^{\mu }-%
\overrightarrow{a_{1}}^{\mu })\cdot\overrightarrow{t}$ and $L^{\mu }=(\omega
^{\mu }+f_{1}^{\mu })t^{0}+(\overrightarrow{\rho }^{\mu }+\overrightarrow{%
a_{1}}^{\mu })\cdot\overrightarrow{t}$ ($\overrightarrow{t}=\frac{1}{2}%
\overrightarrow{\tau },$ where $\overrightarrow{\tau }$ are the Pauli
matrices and $t^{0}=\frac{1}{2}1_{2}$). The corresponding Lagrangian
describing only mesons reads%
\begin{eqnarray}
\mathcal{L}_{mes}& =&\mathrm{Tr}\left[ (D_{\mu }\Phi )^{\dagger }(D^{\mu
}\Phi )-m^{2}\Phi ^{\dagger }\Phi -\lambda _{2}\left( \Phi ^{\dagger }\Phi
\right) ^{2}\right] -\lambda _{1}\left( \mathrm{Tr}[\Phi ^{\dagger }\Phi
]\right) ^{2}+c\,(\det \Phi ^{\dagger }+\det \Phi ) \nonumber \\
& +&\mathrm{Tr}[H(\Phi ^{\dagger }+\Phi
)]-\frac{1}{4}\mathrm{Tr}\left[ (L^{\mu \nu })^{2}+(R^{\mu \nu })^{2}\right]
+\frac{m_{1}^{2}}{2}\mathrm{Tr}\left[ (L^{\mu })^{2}+(R^{\mu })^{2}%
\right]   \nonumber \\
& +&h_{2}Tr(\Phi ^{\dagger }L_{\mu }L^{\mu }\Phi +\Phi R_{\mu }R^{\mu }\Phi
^{\dagger })+h_{3}Tr(\Phi R_{\mu }\Phi ^{\dagger }L^{\mu })+\mathcal{L}_{three}+\mathcal{L}_{four} \; ,  \nonumber \\ \label{meslag}
\end{eqnarray}
where $D^{\mu }\Phi =\partial ^{\mu }+ig_{1}(\Phi R^{\mu }-L^{\mu }\Phi )$
and $L^{\mu \nu }=\partial ^{\mu }L^{\nu }-\partial ^{\nu }L^{\mu }$, $%
R^{\mu \nu }=\partial ^{\mu }R^{\nu }-\partial ^{\nu }R^{\mu }$ are the
field strength tensors of the (axial-)vector fields. $\mathcal{L}_{three}$ and $\mathcal{L}_{four}$
describe 3- and 4-particle interactions of (axial-)vector fields, which are
irrelevant for this work, see Ref.\ \cite{Denisnew}.

The baryon sector involves the baryon doublets $\Psi _{1}$ and $\Psi _{2},$
where $\Psi _{1}$ has positive parity and $\Psi _{2}$ negative parity. In the so-called mirror assignment, $\Psi _{1}$ and $\Psi _{2}$ transform in the
opposite way under chiral symmetry, namely: 
\begin{eqnarray}
\Psi _{1R}& \longrightarrow U_{R}\Psi _{1R},\overline{\Psi }%
_{1R}\longrightarrow \overline{\Psi }_{1R}U_{R}^{\dagger }\;, \;\;\; 
\Psi _{2R}& \longrightarrow U_{L}\Psi _{2R},\overline{\Psi }%
_{2R}\longrightarrow \overline{\Psi }_{2R}U_{L}^{\dagger }\;,  \label{mirror}
\end{eqnarray}%
and similarly for the left-handed fields. Such field transformations allow
to write down a baryonic Lagrangian with a chirally invariant mass term for
the fermions, parametrized by $m_{0}$:
\begin{eqnarray}
{\cal L}_{bar}&=& \overline{\Psi }_{1L}i\gamma _{\mu }D_{1L}^{\mu }\Psi
_{1L}+\overline{\Psi }_{1R}i\gamma _{\mu }D_{1R}^{\mu }\Psi _{1R}+\overline{%
\Psi }_{2L}i\gamma _{\mu }D_{2R}^{\mu }\Psi _{2L}+\overline{\Psi }%
_{2R}i\gamma _{\mu }D_{2L}^{\mu }\Psi _{2R}  \nonumber \\
& -&\widehat{g}_{1}\left( \overline{\Psi }_{1L}\Phi \Psi _{1R}\ +\overline{%
\Psi }_{1R}\Phi ^{\dagger }\Psi _{1L}\right) -\widehat{g}_{2}\left( 
\overline{\Psi }_{2L}\Phi ^{\dagger }\Psi _{2R}\ +\overline{\Psi }_{2R}\Phi
\Psi _{2L}\right)  \nonumber \\
& -&m_{0}(\overline{\Psi }_{1L}\Psi _{2R}-\overline{\Psi }_{1R}\Psi _{2L}-%
\overline{\Psi }_{2L}\Psi _{1R}+\overline{\Psi }_{2R}\Psi _{1L})\;,
\label{nucl lagra}
\end{eqnarray}%
where $D_{1R}^{\mu }=\partial ^{\mu }-ic_{1}R^{\mu }$, $D_{1L}^{\mu
}=\partial ^{\mu }-ic_{1}L^{\mu }$ and $D_{2R}^{\mu }=\partial ^{\mu
}-ic_{2}R^{\mu }$, $D_{2L}^{\mu }=\partial ^{\mu }-ic_{2}L^{\mu }$ are the
covariant derivatives for the nucleonic fields. The coupling constants $%
\widehat{g}_{1}$ and $\widehat{g}_{2}$ parametrize the interaction of the
baryonic fields with scalar and pseudoscalar mesons and $\varphi =$ $%
\left\langle 0\left\vert \sigma \right\vert 0\right\rangle =Zf_{\pi }$ is
the chiral condensate emerging upon spontaneous chiral symmetry breaking in
the mesonic sector. The parameter $f_{\pi }=92.4$ MeV is the pion decay
constant and $Z$ is the wavefunction renormalization constant of the pseudoscalar
fields \cite{Strueber}.

The term proportional to $m_{0}$ generates also a mixing between
the fields $\Psi _{1}$ and $\Psi _{2}.$ The physical fields $N$ and $N^{\ast
},$ referring to the nucleon and to its chiral partner, arise by
diagonalizing the baryonic part of the Lagrangian. As a result we have: 
\begin{equation}
\left( 
\begin{array}{c}
N \\ 
N^{\ast }%
\end{array}%
\right) =\widehat{M}\left( 
\begin{array}{c}
\Psi _{1} \\ 
\Psi _{2}%
\end{array}%
\right) =\frac{1}{\sqrt{2\cosh \delta }}\left( 
\begin{array}{cc}
e^{\delta /2} & \gamma _{5}e^{-\delta /2} \\ 
\gamma _{5}e^{-\delta /2} & -e^{\delta /2}%
\end{array}%
\right) \left( 
\begin{array}{c}
\Psi _{1} \\ 
\Psi _{2}%
\end{array}%
\right) .  \label{mixing}
\end{equation}%
The masses of the nucleon and its partner are obtained upon diagonalizing
the mass matrix $\widehat{M}$:

\begin{equation}
m_{N,N^{\ast }}=\frac{1}{2}\sqrt{4m_{0}^{2}+(\widehat{g}_{1}+\widehat{g}%
_{2})^{2}\varphi ^{2}}\pm \frac{(\widehat{g}_{1}-\widehat{g}_{2})\varphi }{2}%
\;,  \label{nuclmasses}
\end{equation}%
i.e., the nucleon mass is not only generated by the chiral
condensate $\varphi $ but also by $m_{0}$.
The parameter $\delta $ in Eq.\ (\ref{mixing}) is related to the masses
and the parameter $m_{0}$ by the expression: $\delta =\mathrm{Arcosh}\left[ \frac{m_{N}+m_{N^{\ast }}}{2m_{0}}%
\right] \;. 
$%
Let us consider two important limiting cases. (i) When $\delta \rightarrow \infty $, corresponding to $m_{0}\rightarrow 0,$
no mixing is present and $N=\Psi _{1},$ $N^{\ast }=\Psi _{2}.$ In this case $%
m_{N}=\widehat{g}_{1}\varphi /2$ and $m_{N^{\ast }}=\widehat{g}_{2}\varphi
/2 $, thus the nucleon mass is generated solely by the chiral condensate as
in the linear sigma model. (ii) In the chirally restored phase where $\varphi \rightarrow 0$, one has
mass degeneracy $m_{N}=m_{N^{\ast }}=m_{0}.$ When chiral symmetry is broken, 
$\varphi \neq 0$, a splitting is generated. By choosing $0<$ $\widehat{g}%
_{1}<\widehat{g}_{2}$ the inequality $m_{N}<m_{N^{\ast }}$ is fulfilled.

Note also that, when $g_{1}=c_{1}=c_{2}$ and $h_{1}=h_{2}=h_{3}=0$, the chiral symmetry
becomes local \cite{Ko,Pisarski}. The corresponding model has been studied
in Ref. \cite{Wilms}. It was not possible to make a clear-cut prediction as
to whether the mass of the nucleon is dominantly generated by the chiral
condensate or by mixing with its chiral partner. In addition to this, the
description of the mesonic decays was not correct in a locally symmetric
framework as shown in Ref.\ \cite{Parganlija}. Also, the expression for the axial
charge reads $g_{A}^{N}=\frac{\tanh \delta }{Z^{2}}$. Since $\left\vert
\tanh \delta \right\vert <1$ for all $\delta $ and $Z>1$, we obtain $%
g_{A}^{N}<1$, at odds with the experimental value $g_{A}^{N}=1.267\pm 0.004$.
Thus, in the context of local symmetry one is obliged to introduce terms of
higher order such as the Weinberg-Tomozawa one. As a final remark, note that by setting $Z=1$ (which in turn means $g_{1}=0$) the vector mesons and
axial-vector mesons drop out and only the (pseudo-)scalar
and nucleonic terms survive in the Lagrangian. Then, for $g_{A}^{N^{\ast }}$a value much larger
than $0.5$ is predicted, which is in disagreement with lattice data \cite%
{Takahashi}.

In Ref.\ \cite{zschiesche} a large value of the parameter $m_{0}$ ($\sim 800$
MeV) is claimed to be needed for a correct description of nuclear matter
properties, thus pointing to a small contribution of the chiral condensate
to the nucleon mass. Validating this claim through the evaluation of
pion-nucleon scattering at zero temperature and density is the subject of the
present paper.

The expressions for the axial coupling constants of the nucleon and the
partner are given by:%
\begin{eqnarray}
g_{A}^{N} =\frac{e^{\delta }}{2\cosh \delta }g_{A}^{(1)}+\frac{e^{-\delta }%
}{2\cosh \delta }g_{A}^{(2)}, \; \;
g_{A}^{N^{\ast }} =-\frac{e^{-\delta }}{2\cosh \delta }g_{A}^{(1)}+\frac{%
e^{\delta }}{2\cosh \delta }g_{A}^{(2)} \;, \label{ga}
\end{eqnarray}%
where 
\begin{equation}
g_{A}^{(1)}=1-\frac{c_{1}}{g_{1}}\left( 1-\frac{1}{Z^{2}}\right) ,\; \;
g_{A}^{(2)}=-1+\frac{c_{2}}{g_{1}}\left( 1-\frac{1}{Z^{2}}\right)
\end{equation}%
refer to the axial coupling constants of the bare, unmixed fields $\Psi _{1}$
and $\Psi _{2}.$ Note that when $\delta \rightarrow \infty $ one has $%
g_{A}^{N}=g_{A}^{(1)}$ and $g_{A}^{N^{\ast }}=g_{A}^{(2)}.$ Also, when $%
c_{1}=c_{2}=0$ (or $Z=1$) we obtain the results of
Ref.\ \cite{DeTar:1988kn}: $g_{A}^{N}=\tanh \delta $ and $g_{A}^{N^{\ast
}}=-\tanh \delta ,$ which in the limit $\delta \rightarrow \infty $ reduces
to $g_{A}^{N}=1$ and $g_{A}^{N^{\ast }}=-1.$ However, in our model the
interaction with the (axial-)vector mesons generates additional
contributions to $g_{A}^{N}$ and $g_{A}^{N^{\ast }}$, which are fixed via
the experimental result for $g_{A}^{N}$ and the lattice result for $%
g_{A}^{N^{\ast }},$ see the next section.
From the Lagrangian (\ref{nucl lagra}) one can compute the decay $N^{\ast }\rightarrow N\pi$ and the scattering amplitudes $a_{0}^{(\pm )}$ \cite{Gallas}.

\section{ Results and discussion}

We consider three possible assignments for the partner of the nucleon. (1)
The resonance $N^{\ast }(1535)$, with mass $M_{N^{\ast }(1535)}=1535$ MeV
and $\Gamma _{N^{\ast }(1535)\rightarrow N\pi }=(67.5\pm 23.6)$ MeV, which
-- being the lightest baryonic resonance with the correct quantum numbers --
surely represents one of the viable and highly discussed candidates for the
nucleon partner. (2) The resonance $N^{\ast }(1650),$ with a mass lying just
above, $M_{N^{\ast }(1650)}=1650$ MeV and $\Gamma _{N^{\ast
}(1650)\rightarrow N\pi }=(92.5\pm 37.5)$ MeV. (3) A $speculative$ candidate 
$N^{\ast }(1200)$ with a mass $M_{N^{\ast }(1200)}\sim 1200$ MeV and a very
broad width $\Gamma _{N^{\ast }(1650)\rightarrow N\pi }\gtrsim 800$ MeV,
such to have avoided experimental detection up to now \cite{zschiesche}.

For all these scenarios, we want to determine the values of the parameters $%
c_{1},$ $c_{2}$, and $m_{0}.$ Beyond the width, which is different in the three
cases mentioned above, we also use $g_{A}^{N^{\ast }}=0.2\pm 0.3$, as
predicted by lattice QCD \cite{Takahashi}, and $g_{A}^{N}=1.26$. We repeat
the evaluation for different values of $Z.$ Remarkably, $m_{0}$ does not
depend on $Z$.

\begin{figure}[h]
\centering
\includegraphics[scale = 0.64]{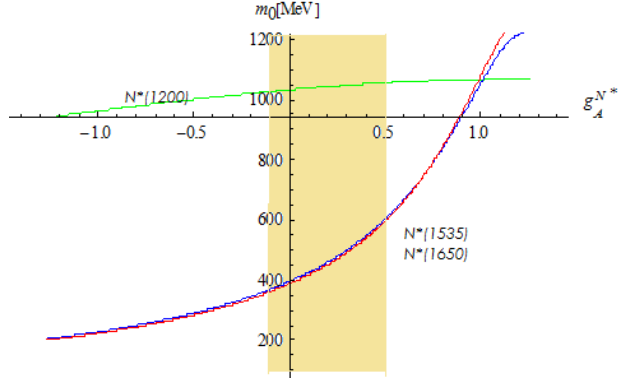}
\caption{$m_{0}$ as a function of $g_{A}^{N^{\ast }}$. }
\label{allem0_glob}
\end{figure}

Figure \ref{allem0_glob} shows the mass parameter $m_{0}$ as a function of the
axial coupling constant of $N^{\ast }$ for different masses of $N^{\ast }$.
For the range of $g_{A}^{N^{\ast }}$ given by Ref.\ \cite{Takahashi}, $%
m_{0}=300-600 $ MeV for $N^{\ast }(1535)$ and $N^{\ast }(1650)$, meaning
that half of the nucleonic mass survives in the chirally restored phase.
On the contrary, for $N^{\ast }(1200)$ the value for $m_{0}$ lies above $%
1000 $ MeV. This result suggests that the contribution of the chiral
condensate to the nucleonic mass should be negative, which is rather
unnatural. We can then discard the possibility that a
hypothetical, not yet discovered\ $N^{\ast }(1200)$ is the chiral partner of
the nucleon.

According to Ref.\ \cite{Glozman}, when the fields $N$ and $N^{\ast }$ belong
to a parity doublet, $g_{A}^{N}\sim 1$ and $%
g_{A}^{N^{\ast }}\sim -1$. Then, in Ref. \cite{Glozman} the lattice
result of \cite{Takahashi} is used against the identification of $N^{\ast
}(1535)$ as the partner of the nucleon. However, within our model we can 
{\em still\/} accommodate $N^{\ast}(1535)$ (or also $N^{\ast }(1650)$) as the
partner of the nucleon. The small value of $g_{A}^{N^{\ast }}$ arises 
because of interactions of the partner with the axial-vector mesons.

%\FRAME{ftbpF}{5.0548in}{3.1488in}{0pt}{}{}{allem0_glob.bmp}{\special%
%{language "Scientific Word";type "GRAPHIC";maintain-aspect-ratio
%TRUE;display "USEDEF";valid_file "F";width 5.0548in;height 3.1488in;depth
%0pt;original-width 5.0004in;original-height 3.1038in;cropleft "0";croptop
%"1";cropright "1";cropbottom "0";filename '../Plots Diss/Plots globales
%Modell/allem0_glob.bmp';file-properties "XNPEU";}}

\begin{figure}[h]
\centering
\includegraphics[scale = 0.64]{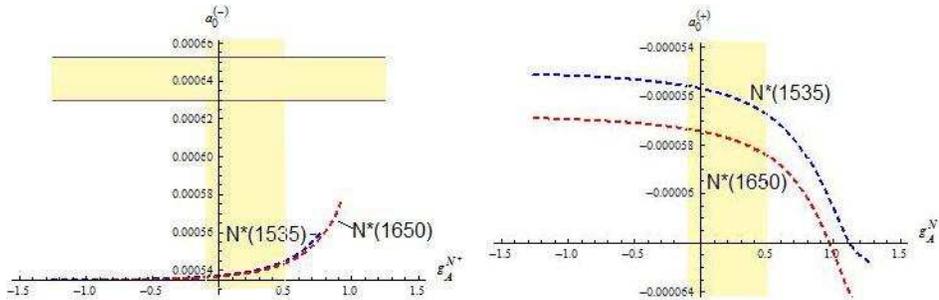}
\caption{The scattering lengths (a) $a_{0}^{(-)}$ and (b) $a_{0}^{(+)}$ as a function of $g_{A}^{N^{\ast }}$.}
\label{Figure}
\end{figure}

In Figure \ref{Figure} we plot the isospin-odd and isospin-even
scattering lengths as a function of the axial charge $g_{A}^{N^{\ast }}$for $%
Z=1.5$. A comparison of our results to the experimental data on $\pi N$
scattering lengths, as measured in Ref.\ \cite{schroder} by precision X-ray
experiments on pionic hydrogen and pionic deuterium, yields the following:
(a) the isospin-odd scattering length $%
a_{0}^{(-)}$ is close to the experimental range, but we expect an even better
result with the introduction of the $\Delta $ resonance.
(b) The isospin-even scattering length $a_{0}^{(+)}$ is an order of magnitude smaller than the experimental band $a_{0,\exp }^{(+)}=(-8.85783\pm
7.16)10^{-6}$ MeV. The reason for this is the strong dependence of $%
a_{0}^{(+)}$ on the scalar mesons. Here we use $m_{\sigma }=1370$ MeV. A
smaller mass of the sigma meson may be favored, however this result seems to
be excluded \cite{Denisnew}. 
%\bigskip \FRAME{ftbpF}{%
%6.7499in}{2.1266in}{0pt}{}{}{a0plus und a0minus.jpg}{\special{language
%"Scientific Word";type "GRAPHIC";maintain-aspect-ratio TRUE;display
%"USEDEF";valid_file "F";width 6.7499in;height 2.1266in;depth
%0pt;original-width 8.5417in;original-height 2.6671in;cropleft "0";croptop
%"1";cropright "1";cropbottom "0";filename 'a0plus und
%a0minus.jpg';file-properties "XNPEU";}}

\section{Summary and outlook}

We have computed the pion-nucleon scattering lengths at tree-level in the
framework of a globally symmetric linear sigma model with parity-doubled
nucleons. Within the mirror assignment the mass of the nucleon originates
only partially from the chiral condensate, but also from the mass parameter $%
m_{0}$. Using the lattice results of Ref.\ \cite{Takahashi} we find that $m_{0}\simeq300-600$ MeV. Approximately half of the nucleon mass survives in the chirally restored
phase.
The isospin-odd scattering length lies close to the experimental band, but
could be improved in further studies.
The isospin-even scattering length is too small: future inclusion of a light tetraquark state gives rise to a large contribution, and thus is expected to improve the results. We also plan to extend our model by including the $\Delta $
resonance, necessary to correctly reproduce the p-wave scattering
lengths \cite{Procura} and to evaluate the radiative
 $\eta $-photoproduction.

\section*{Acknowledgements}

The authors thank T.\ Kunihiro, T. Takahashi and D. Parganlija for useful discussions.

\end{document}